\documentclass[12pt]{article}
\usepackage{epsfig}     

\newcommand{\nn}{\nonumber}

\newcommand\be{\begin{equation}}
\newcommand\ee{\end{equation}}
\newcommand\bea{\begin{eqnarray}}
\newcommand\eea{\end{eqnarray}}

\newcommand{\phd}{\phi^{\dagger}}

\def\ma[#1,#2,#3,#4]  {{\left( \matrix{ #1  & #2 \cr
                                        #3  & #4 \cr } \right)}}

\begin{document}

\title{
{\vspace{-0cm} \normalsize
\hfill \parbox{40mm}{CERN/TH-98-272}\\
\hfill \parbox{40mm}{MPI-PhT/98-66}}\\[25mm]
The PHMC algorithm for \\simulations of dynamical fermions:\\
II - Performance analysis}
\author{
Roberto  Frezzotti$^1$  
and Karl Jansen$^{2,}$\footnote{Heisenberg Foundation Fellow}  
\\
{\footnotesize
$^1$ Max-Planck-Institut f\"ur Physik, F\"ohringer Ring 6, D-80805
     Munic, Germany
}
\\
{\footnotesize 
$^2$ CERN, 1211 Gen\`eve 23, Switzerland  
}
}
\date{\today}
\maketitle

\begin{abstract}

We compare the performance of the PHMC algorithm with the
one of the HMC algorithm in practical simulations of lattice QCD.
We show that the PHMC algorithm can lead to an acceleration
of numerical simulations. It is demonstrated that the
PHMC algorithm generates configurations carrying small
isolated eigenvalues of the lattice Dirac operator
and hence leads to a sampling
of configuration space that is different from that    
of the HMC algorithm. 

\end{abstract}

\section*{Introduction}

In this paper we continue our discussion of the Polynomial
Hybrid Monte Carlo (PHMC) algorithm \cite{frezzi,hitech}.
This algorithm, designed for simulations of models containing
fermionic degrees of freedom, is based on the idea \cite{takaishi} 
of combining
the Hybrid Monte Carlo (HMC) algorithm \cite{hmc} with the multiboson technique
\cite{martin}. 
In the PHMC algorithm the update part relies on an approximation of the exact
fermion action to be simulated. The error induced by this approximation is 
corrected for 
by a reweighing technique, which introduces a correction factor
taken into account 
in the sample average of the observables.

In this paper we will present our results 
concerning the dynamical behaviour of 
the PHMC algorithm in practice. On the quantitative level we will
compare its performance with the one of the HMC algorithm.
Our numerical tests have
been done 
in the Schr\"odinger functional set up \cite{SF:LNWW,sint,paper3},  
on small and moderately large physical volumes but at almost
vanishing quark mass, which is feasible when using
Schr\"odinger functional boundary conditions. 
We remark that since we are working at tiny values
of the quark mass, the condition number of the fermion matrix 
employed in our simulations becomes ${\rm O}(2000)$. 

We will demonstrate that the PHMC algorithm 
samples configuration space differently from
the HMC algorithm. In particular, using the PHMC algorithm,
gauge configurations with very small eigenvalues 
of the lattice Dirac operator can be reached.
Consequences of this behaviour on the results for physical observables
are discussed. 

We assume that the reader is familiar with refs.~\cite{frezzi,hitech}.
In particular, in the latter reference we have discussed a number of relevant technical 
aspects, which lay the ground for the present performance analysis.

\section{Numerical simulations with \\ the PHMC algorithm}

In order to make the paper reasonably self-contained,
we summarize here some features of the PHMC algorithm. 
We remark that 
throughout the paper we will use 
${\rm O}(a)$-improved Wilson fermions.                                

\subsection{Ingredients of the PHMC algorithm}

Denoting the lattice gauge link from $x$ to $x+a\hat{\mu}$
by $U_{\mu}(x) \in SU(3)$ and a 
gauge field configuration by $U$, 
the expectation value of any gauge invariant observable ${\cal O}=
{\cal O}[U]$, in full QCD with $n_f =2$ degenerate flavours,
may be written as
\be
\langle {\cal O} \rangle = {\cal Z}^{-1}
                               \left[ \int {\cal D}U
e^{- S_g[U]} \mbox{det}(Q^2[U]) {\cal O}[U] \right]\; ,
\ee
where $S_g$ is the standard plaquette action for the pure gauge
sector and $Q$ is the Dirac operator for 
${\rm O}(a)$-improved Wilson
fermions multiplied by $\gamma_5$ (see below). 

The PHMC algorithm makes use of a polynomial approximation of $(Q^2)^{-1}$.
The polynomial in a real variable $s$ and having 
degree $n$ is denoted by
$P_{n,\epsilon}(s)$ and constructed such that it approximates 
$s^{-1}$ in the range $0 < \epsilon
< s <1$ with a relative fit error bounded by 
\begin{equation} \label{delta}
\delta =
2 \left(\frac{1-\sqrt{\epsilon}}{1+\sqrt{\epsilon}}\right)^{n+1}\; .
\end{equation}
We choose the normalization of the Dirac operator
such that the highest eigenvalue of $Q^2$ is smaller than $1$ and 
write the corresponding polynomial in $Q^2$, $P_{n,\epsilon}(Q^2)$, in a factorized
form:
\be \label{fact_form}
P_{n,\epsilon}(Q^2) =
C_{n,\epsilon} \prod_{k=1}^n (Q-r_k^*)(Q-r_k)\; , 
\ee
where $ C_{n,\epsilon} $ is a positive constant, $r_k$ is determined by 
$\sqrt{z_k}$ (see \cite{hitech} for the exact relation),  
and the $z_k$'s are the complex roots of
$P_{n,\epsilon}(s)$.
We note that special care has to be taken concerning the precise ordering
of the factors in eq.~(\ref{fact_form})  
in order to avoid problems with rounding errors \cite{rootorder}.

The full QCD ($n_f=2$) partition function may
now be represented as 
\bea
{\cal Z} &
 = & \int {\cal D}U{\cal D}\phd{\cal D}\phi
{\cal D}\eta^\dagger{\cal D}\eta
  W e^{- (S_g + S_{P}+\eta^\dagger\eta) } \nn \\
  S_{P} & = & S_{P}[U,\phi]=\phd P_{n,\epsilon}(Q^2[U]) \phi
     \label{S_g_P}
\eea
by introducing the auxiliary pseudofermion fields (i.e. boson fields with
spin and colour indices) $\phi$, $\eta$ and the correction
factor $W=W[\eta,U]$:
\be \label{W_def}
W= \exp\left\{\eta^\dagger (1-[Q^2\cdot P_{n,\epsilon}(Q^2)]^{-1}) \eta \right\}
 \; .
\ee

Each evaluation of $W$ requires a trivial
Gaussian ``update'' of the $\eta$-field and the solution of
the system $[Q^2 P_{n,\epsilon}(Q^2)]\chi = \eta$.
In practice it turned out to be 
useful to generate the $\eta$-fields $N_{\rm corr}$ times for each given 
gauge field configuration. 
Denoting averages evaluated with the effective action $S_g + S_{P}+\eta^\dagger\eta$
as $\langle \dots \rangle_P$, the exact averages denoted as
$\langle \dots \rangle$ are obtained by reweighing with $W$
\be
\langle {\cal O} \rangle = \langle W \rangle_P^{-1}
\langle {\cal O}W \rangle_P  \label{true_ave} \; .
\ee

In \cite{frezzi} we presented some tests of the PHMC algorithm
for non-improved Wilson fermions. 
In this paper, 
we extend these tests to the
case of ${\rm O}(a)$-improved actions. 
With respect to
the non-improved case this amounts to adding the so--called ``clover''
term \cite{clover} to the lattice Dirac operator, as specified below. The modifications
in the PHMC algorithm
induced by this extra term in the action are completely analogous to the ones
needed for the standard HMC algorithm and our implementation of the PHMC
algorithm for ${\rm O}(a)$-improved fermions follows closely 
the procedure described in \cite{janliu} for the HMC algorithm.

For the actual simulations
we consider hypercubic space-time lattices with 
lattice spacing $a$ and
size $L^3\times T$. With the lattice spacing set to unity from now on, the
points on the lattice have integer coordinates $(x_0,x_1,x_2,x_3)$, which
are in the range $0\le x_0 \le T;0\le x_i < L$.
The gauge and the fermion fields 
obey Schr\"odinger functional
boundary conditions as used in \cite{jansom} and detailed in \cite{SF:LNWW,sint,paper3}. 
The matrix defining the fermion action will be denoted 
by $Q$:
\begin{eqnarray} 
\label{qmatrix}
Q(U)_{xy} \!\!\!&=& \!\!\!\frac{c_0}{c_M}\gamma_5 [
(1+\sum_{\mu\nu}
[{i \over 2}c_{\rm sw}\kappa\sigma_{\mu\nu}{\cal F}_{\mu\nu}(x)])\delta_{x,y}
 \nonumber \\
&-&\kappa\sum_{\mu} \{
   (1-\gamma_{\mu})U_{\mu}(x)\delta_{x+\mu,y} +
(1+\gamma_{\mu})U^{\dagger}_{\mu}(x-\mu)\delta_{x-\mu,y}\}]  \;\;,
\end{eqnarray} 
where $\kappa$ is the hopping parameter and $c_{\rm sw}$ the improvement coefficient.           
The constant 
$c_M$ serves to optimize 
the simulation algorithm and $c_0=(1+8\kappa)^{-1}$.
For further unexplained notations we refer to refs.~\cite{hitech,rootorder}. 

In order to speed up the Monte Carlo simulation,
 not the original matrix $Q$ 
but an even-odd preconditioned \cite{precond} matrix $\hat{Q}$ is used. 
We expect the algorithm to be working equally well by using different preconditioning
techniques like SSOR \cite{ssor}. 
Let us rewrite the matrix $Q$ in eq.~(\ref{qmatrix}) as
\begin{equation} 
Q \equiv \frac{c_0}{c_M} \gamma_5  \left( \begin{array}{cc}
                1+T_{ee} & M_{eo} \\
                M_{oe} & 1+T_{oo} \\
                \end{array} \right)\;\;,
\end{equation} 
where we introduce the matrix $T_{ee}$ ($T_{oo}$) on the
even (odd) sites as
\begin{equation} \label{eq:t}
(T)_{xa\alpha,yb\beta} =
\sum_{\mu\nu}[{i \over 2}c_{\rm sw}\kappa\sigma^{\alpha\beta}_{\mu\nu}
{\cal F}^{ab}_{\mu\nu}(x) \delta_{xy}]\;\;.
\end{equation}
The off-diagonal parts $M_{eo}$ and $M_{oe}$
connect the even with odd and the odd with even lattice
sites, respectively.        
Preconditioning is now realized 
by writing the determinant of $Q$,
apart from an irrelevant constant factor, as
\begin{eqnarray} \label{q_oo_matrix}
\det(Q)&\propto&\det(1+T_{ee})\det\hat{Q}
\nonumber \\
\hat{Q}&=&\frac{\hat{c}_0}{c_M} \gamma_5
(1 + T_{oo} - M_{oe}(1+T_{ee})^{-1}M_{eo}) \;\;.
\end{eqnarray} 
The constant factor
$\hat{c}_0$ is given by $\hat{c}_0=1/(1+64\kappa^2)$,
and the constant $c_M$ 
is chosen such that the eigenvalues of $\hat{Q}$ are
well within the interval $[-1,1]$. 
Since for the simulation algorithms the eigenvalues have to be positive,
we finally work with the matrix $\hat{Q}^2$. We note that in the
case $c_{\rm sw} \neq 0$ the PHMC algorithm makes use of the polynomial 
approximation $P_{n,\epsilon}(\hat{Q}^2) \simeq 1/(\hat{Q}^2)$ 
only to simulate $\det\hat{Q}^2$, while the term $\det(1+T_{ee})^2$
is treated exactly. The correction factor 
therefore accounts only for the missing contribution, i.e. 
$\det\hat{Q}^2 P_{n,\epsilon}(\hat{Q}^2)$, to the partition function.

\subsection{The simulations} 

An important question is how the parameters of the polynomial $P_{n,\epsilon}$ are
to be chosen. Following ref.~\cite{hitech}, 
a practical recipe for the choice of $\epsilon$ and $n$ may be given by 
\be \label{epsilon}
\epsilon \simeq
2 \frac{\langle \lambda_{\rm min}(\hat{Q}^2) \rangle}{\langle \lambda_{\rm max}(\hat{Q}^2) \rangle}
\ee
and the value of $n$ is set such that $\delta \simeq 0.01$ (see eq.~(\ref{delta})). 
In eq.~(\ref{epsilon}) 
$\lambda_{\rm min}(\hat{Q}^2)$ and $\lambda_{\rm max}(\hat{Q}^2)$ denote the lowest
and the highest eigenvalues of $\hat{Q}^2$, respectively. Our experience suggests
that only a poor knowledge of the value of the average condition
number
$k=\langle \lambda_{\rm max}(\hat{Q}^2) / \lambda_{\rm min}(\hat{Q}^2) \rangle$
for the specific run parameters is needed. We remark that 
$k\approx \langle \lambda_{\rm max}(Q^2)\rangle/\langle\lambda_{\rm min}(Q^2) \rangle$. 
An estimate of $k$ can be obtained, e.g. in 
the thermalization phase of the simulation, which may be performed by using either  
the standard HMC algorithm or the PHMC algorithm itself. 
We have also found that
a very good and decisive check about the quality of the chosen polynomial
approximation can be performed by monitoring the fluctuations of the
correction factor $W$: using too poor a polynomial approximation to
$\hat{Q}^{-2}$ gives rise to large fluctuations of $W$, and consequently 
large fluctuations of many
reweighted observables (eq.~(\ref{true_ave})), which can be detected after
a few trajectories. 

Another remark 
\footnote{We are grateful to A.D. Kennedy for this argument.}
concerns the dependence of the approximation
on the volume $V$:
the difference of the actions $\Delta_S = S_{\rm HMC} - S_{\rm PHMC}$ is
asymptotically $\Delta_S = V \; C_S \; \exp(-2\sqrt{\epsilon}n)$, with $C_S$ 
some proportionality constant. Since 
$\epsilon$ is fixed by the condition of eq.~(\ref{epsilon}) we find, if we also 
want to keep $\Delta_S$ fixed, that 
$n\simeq - \; (\log \Delta_S - \log V -\log C_S)/2\sqrt{\epsilon} $.
We see that the explicit volume dependence in $n$ is rather weak 
in comparison with the (power--like) volume dependence induced by the way we 
choose $\epsilon$, 
following the criterion of eq.~(\ref{epsilon}). We therefore expect that
the PHMC algorithm will also work efficiently in the case of large volumes, 
while keeping the value of $\delta$, eq.~(\ref{delta}), about $0.01$.

The numerical tests 
are performed on $8^3\times 16$ lattices using 
the massively parallel Alenia Quadrics (APE) machines. 
Simulation parameters were chosen to be 
\begin{eqnarray}  \label{runpar_clover}
\beta=6.8 & \quad , \quad \kappa = 0.1343 & \quad , \quad c_{\rm sw} = 1.4251 \\  
\beta=5.4 & \quad , \quad \kappa = 0.1379 & \quad , \quad c_{\rm sw} = 1.7275 \; .
\end{eqnarray}
These parameter values correspond to those used 
in  simulations
to determine the values of $c_{\rm sw}$ non-perturbatively \cite{jansom}. 

All tests described below were performed on the APE machines by 
running $N_{\rm rep}$ replica in parallel, with $N_{\rm rep}$ set
to 32 or 16. 
Since the $N_{\rm rep}$ replica were independently thermalized,
the data from the different replica are statistically independent
from each other. This allows for a reliable
error analysis, provided that for each replicum the statistics
is several times larger than the integrated autocorrelation time of the 
observable considered. 
We determined our statistical errors for the observables,  given below, 
from the variance of the
$N_{\rm rep}$ data obtained from running in parallel. 
We checked that the results were consistent with those obtained from 
a jack-knife procedure combined with binning. We refer to
\cite{hitech} for further details. 
 
It is also possible to divide the $N_{\rm rep}$ system replica 
into 2 sets of $N_{\rm rep}/2$  replica and
analyse each of these two sets of data (a and b) separately.
This gives two errors $\Delta_a$ and $\Delta_b$, each of them obtained with
half the statistics of the full run. By rescaling
$\Delta_a$ and $\Delta_b$ by $\sqrt{2}$, we can obtain an
estimate of the error on the error, which in turn gives a measure of the error
on the integrated autocorrelation time. 
This way of determining the error on the error yields values that are compatible
with the estimate \cite{hitech,jansom} of the relative error on the error 
given by $(2N_{\rm rep})^{-1/2}$,  or, in the case of 
a binning analysis leading to $N_{\rm block}$ independent measurements, by
$(2N_{\rm block})^{-1/2}$. In the latter case, of course, the values of $N_{\rm block}$
have to be large enough that a plateau behaviour of the error can be detected.
For a few tables below, 
besides the mean values and the errors (indicated in round brackets),
we also quote the error on the error (indicated in square brackets).

\section{Results at $\beta=6.8$}

In this section we give our results for various quantities 
as computed with the PHMC algorithm and compare with those  
obtained from the HMC algorithm. We will compare bulk quantities
as well as 
quark correlation functions and certain combinations of them.  

We give in table~\ref{tech68} the parameters for both simulation
algorithms. As reported in \cite{jansom}, in the simulations with the
HMC algorithm, we found that sometimes a trajectory was not accepted for
a number of times. The cure was that every $l$-th trajectory the step size 
$\delta\tau$ was changed to a smaller value, and the corresponding number of
molecular dynamics steps, $N_{\rm md}$, was increased to reach a unit
trajectory length. In the actual simulation a value of $l=6$ was chosen
and we give in table~\ref{tech68} the effective values of $\delta\tau$ and
$N_{\rm md}$ built from the normal and the smaller step size. We remark
that this effect, observed in simulations with the HMC algorithm, 
never appeared within the simulations using the PHMC algorithm and 
that the step size was always kept constant there. 
This allowed in particular to run the PHMC algorithm at an           
acceptance rate smaller than the one obtained with the HMC algorithm.

\begin{table*}[hbt]
\caption{The parameters for both simulation algorithms
at $\beta=6.8$. 
We denote by 
Stat the statistics obtained in units of
trajectories and $P_{\rm acc}$ is the acceptance rate.}
\vspace{2mm}
\label{tech68}
\begin{tabular}{llllllll}
\hline
Algorithm & $\delta\tau^{\rm eff}$ & $N_{\rm md}^{\rm eff}$ & $P_{\rm acc}$ & Stat & $\epsilon$ & $n$ & $c_M$  \\
\hline \hline
 HMC        & $0.059$ & $16.6$ & $0.948(8)$ & $2944$ & --  & -- & -- \\
 PHMC       & $0.077$ & $13$ & $0.79(1)$    & $2944$ & $0.0022$ & $62$ & $0.735$  \\
 PHMC$^{*}$ & $0.077$ & $13$ & $0.758(8)$   & $2944$ & $0.0022$ & $54$ & $0.725$  \\
\hline \hline
\end{tabular}
\end{table*}

\begin{table*}[hbt]
\caption{Comparison of bulk quantities as obtained from the two algorithms. 
We use the notation 
PHMC($N_{\rm corr}$) to indicate the values of $N_{\rm corr}$ 
used for the analysis. $N_{\rm corr}=0$ means that the 
correction factor is set to 1. 
In square brackets we give our estimate of the error on the error.}
\vspace{2mm}
\label{tablebulk}
\begin{tabular}{lllll}
\hline
Algorithm & \makebox[1.5cm][r]{$\langle P \rangle$}& $\langle \lambda_{\rm min}(\hat{Q}^2) \rangle $ 
          &   $\langle \lambda_{\rm max}(\hat{Q}^2) \rangle $ & $\langle dS_g/d\eta \rangle$ \\
\hline \hline
 HMC     & $0.673384(53)[7]$ & $0.001150(35)[4]$ & $0.87188(25)[3]$ & $23.1(2.4)[0.3]$  \\
 PHMC(4) & $0.673483(46)[6]$ & $0.001152(42)[5]$ & $0.87164(36)[5]$ & $22.1(2.3)[0.3]$  \\
 PHMC(2) & $0.673496(45)[6]$ & $0.001150(43)[5]$ & $0.87173(39)[5]$ & $22.6(2.3)[0.3]$  \\
 PHMC(1) & $0.673505(48)[6]$ & $0.001141(42)[5]$ & $0.87190(42)[5]$ & $22.6(2.2)[0.3]$  \\
 PHMC(0) & $0.673512(44)[6]$ & $0.001025(46)[6]$ & $0.87177(30)[4]$ & $20.6(2.0)[0.3]$  \\
\hline
PHMC$^*$(2) & $0.673435(66)[8]$ & $0.001117(44)[6]$ & $0.87426(70)[9]$ & $27.1(3.1)[0.4]$  \\
\hline \hline
\end{tabular}
\end{table*}
 
\subsection{Bulk quantities}
 
As bulk quantities we consider the expectation values for
the plaquette $P$, the lowest $\lambda_{\rm min}$ and the largest
$\lambda_{\rm max}$ eigenvalues of $\hat{Q}^2$, and the derivative of the
pure gauge action with respect to the background field,
$dS_g/d\eta$. The latter quantity can be used to define
a running coupling constant in the pure gauge theory
\cite{su3paper}.
 
As table~\ref{tablebulk} shows, we find that, for $N_{\rm corr}>0$ the
values of all bulk quantities are 
completely consistent with the corresponding ones from the HMC run. 
Also the uncorrected (see $N_{\rm corr}=0$) 
values for $\langle \lambda_{\rm max}\rangle$ and 
$\langle dS_g/d\eta \rangle$ are in agreement with the HMC values while, 
perhaps, the ones for 
$\langle P \rangle$ and $\lambda_{\rm min}$ are somewhat off. 
Within the error on the error, also the estimated errors on the
observables are consistent between the PHMC and HMC algorithm.

The prominent exception is $\lambda_{\rm max}$, where  
the error from the PHMC algorithm appears to be substantially larger than the one
from the HMC algorithm. 
Note, however, that the mean value and the error for the uncorrected
value of $\lambda_{\rm max}$ are both consistent with the corresponding 
quantities from the HMC algorithm. In addition, the error decreases when
$N_{\rm corr}$ is increased from 1 to 4. This points towards the interpretation 
that the larger error is just induced by the additional noise appearing through 
the correction factor and that there is no large autocorrelation time hidden
in the PHMC algorithm.
Of course, $\lambda_{\rm max}$ is a pure cut-off quantity and is not expected
to be physically relevant. 



\subsection{Quark correlation functions}

Quark correlation functions are important quantities,  
from which many physical observables can be constructed.
We hence extend our comparison of the algorithms by considering
certain quark correlation functions, which are often used 
in computations with Schr\"odinger functional boundary conditions.
To this end we closely follow ref.~\cite{paper3} and
construct correlation functions
using boundary
quark fields $\zeta$, $\bar{\zeta}$ at Euclidean time $x_0=0$:
%
%
\begin{eqnarray} \label{fafp}
f_A(x_0) & = & - \sum_{\bf y,z} \frac{1}{3} A_0^a(x) \bar{\zeta}
             (0,{\bf y}) \gamma_5\frac{1}{2}\tau^a \zeta (0,{\bf z})
             \nonumber \\
f_P(x_0) & = & - \sum_{\bf y,z} \frac{1}{3} P^a(x) \bar{\zeta}
             (0,{\bf y}) \gamma_5\frac{1}{2}\tau^a \zeta (0,{\bf z})
             \; .
\end{eqnarray}
In eq.~(\ref{fafp}) $A_0^a(x)$ denotes the isovector axial current and 
$P^a(x)$ the corresponding density
\begin{eqnarray} \label{AP}
A_\mu^a & = & \bar{\psi}\gamma_\mu\gamma_5\frac{1}{2}\tau^a\psi \nn \\
P^a & = & \bar{\psi}\gamma_5\frac{1}{2}\tau^a\psi\; , 
\end{eqnarray}
where $\tau^a$ is a Pauli matrix acting on the flavour indices of the quark field. 
 
Analogously one may build $f_A'(T-x_0)$
and $f_P'(T-x_0)$
with boundary quark fields
$\zeta'$, $\bar{\zeta}'$ at $x_0=T$.

We will consider the correlation functions $f_A(x_0)$, $f_P(x_0)$ as well 
as finite differences of them:
\begin{eqnarray} \label{differences}
d_A(x_0) & = &  (\partial^{*}_0 + \partial_0) f_A(x_0)\; ,\; 0 < x_0 < T
             \nonumber \\
D_P(x_0)& = & \partial^{*}_0\partial_0 f_P(x_0)\; ,\; 0 < x_0 < T
             \; .
\end{eqnarray}
In eq.~(\ref{differences}) $\partial_0$ is the lattice forward derivative,
and $\partial^{*}_0$ the lattice backward derivative
\begin{eqnarray} \label{derivatives} 
\partial_0 f(x_0) & = & f(x_0+1)-f(x_0) \nn \\
\partial_0^{*} f(x_0) & = & f(x_0)-f(x_0-1)\; .
\end{eqnarray}

\begin{table*}[hbt]
\caption{Comparison of quark correlation functions as obtained
from the two algorithms. Notations are as in table~\ref{tablebulk}. 
}
\vspace{2mm}
\label{tablecorr}
\begin{tabular}{lllll}
\hline
Algorithm &  $\langle f_A(T/2) \rangle\cdot 10^{-4}$ & $\langle f_P(T/2)\rangle\cdot 10^{-6}$ 
 &  $\langle d_A(T/2) \rangle\cdot 10^{-4}$ & $\langle D_P(T/2) \rangle\cdot 10^{-5}$ \\
\hline \hline
 HMC     & $0.542(39)[5]$  & $0.1072(55)[7]$ & $-0.171(13)[2]$ & $0.2062(60)[8]$  \\
 PHMC(4) & $0.609(58)[7]$  & $0.1074(51)[6]$ & $-0.171(12)[2]$ & $0.1957(55)[7]$  \\
 PHMC(2) & $0.612(60)[8]$  & $0.1075(53)[7]$  & $-0.168(13)[2]$ & $0.1969(56)[7]$  \\
 PHMC(1) & $0.618(60)[8]$  & $0.1081(53)[7]$  & $-0.170(13)[2]$ & $0.1986(56)[7]$  \\
 PHMC(0) & $0.879(144)[18]$ & $0.1300(80)[10]$  & $-0.193(15)[2]$ & $0.1891(83)[10]$  \\
\hline
PHMC$^*$(2) & $0.632(70)[9]$ & $0.1088(56)[7]$  & $-0.173(12)[2]$ & $0.1939(50)[6]$  \\
\hline \hline
\end{tabular}
\end{table*}
 
We compare the results for quark correlation functions as obtained from 
the two algorithms in 
table~\ref{tablecorr}. We take the distance in time to be half the temporal size
of the lattice, i.e. $x_0 = T/2$. 
We can see from table~\ref{tablecorr} that we have again consistent results 
for the mean values of 
$f_P(T/2)$, $d_A(T/2)$ and $D_P(T/2)$, as well as for the corresponding errors. 
However, for $f_A(T/2)$ the error from the PHMC algorithm is substantially
larger than the one from the HMC algorithm. The discrepancy is well outside
the uncertainty of the error as indicated by the error on 
the error given in the square brackets. 
Even more pronounced is the behaviour of the uncorrected value of 
$f_A(T/2)$. The error is a factor of about 4 larger than in the HMC case, and
also the mean value is off. 

\begin{figure}[t]
\vspace{-0mm}
\centerline{ \epsfysize=12.5cm
             \epsfxsize=12.5cm
             \epsfbox{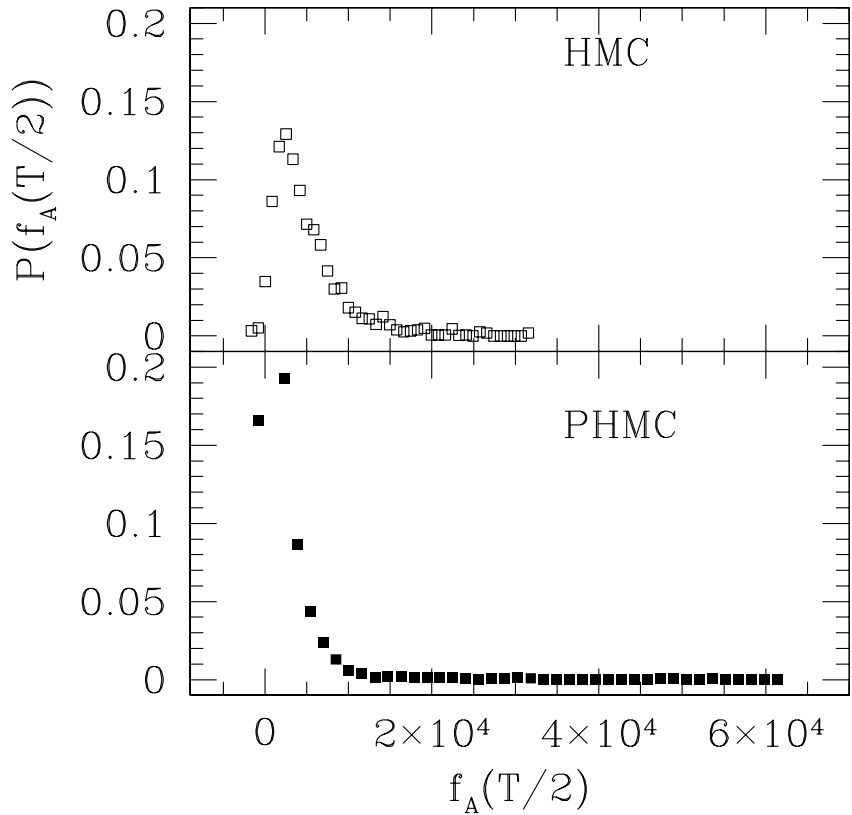}}
\begin{center}
\parbox{12.5cm}{\caption{ \label{fig:fadis}
The distributions of the quark correlation function $f_A(T/2)$   
as obtained from the HMC and the PHMC algorithms at $\beta=6.8$. 
Note that for the PHMC algorithm we plot the uncorrected values 
for $f_A(T/2)$. 
}}
\end{center}
\end{figure}

\begin{figure}
\vspace{-0mm}
\centerline{ \epsfysize=18.5cm
             \epsfxsize=18.5cm
             \epsfbox{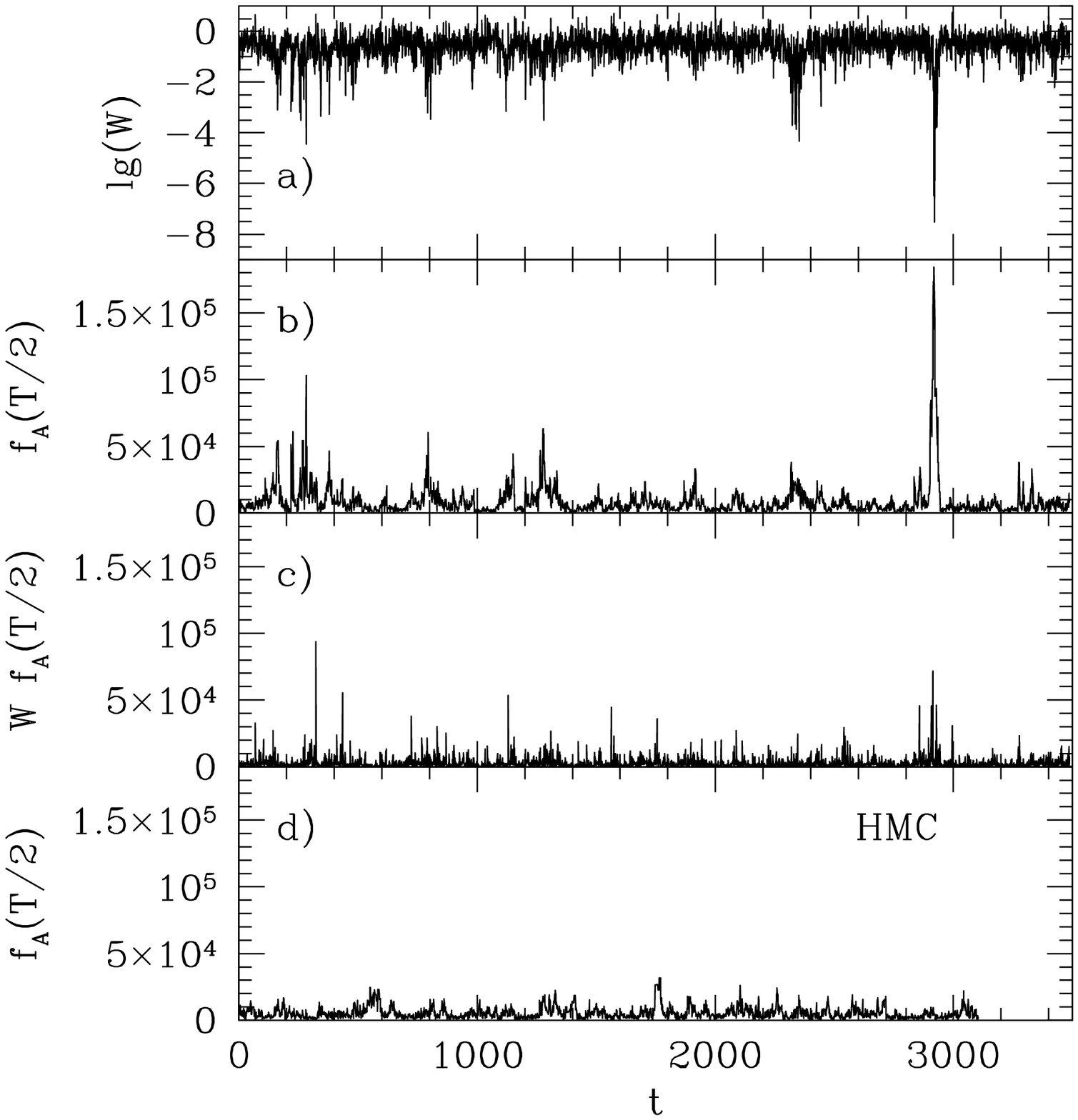}}
\begin{center}
\parbox{12.5cm}{\caption{ \label{fig:allfa}
The Monte Carlo time evolutions for $W$, $f_A(T/2)$, $Wf_A(T/2)$, as obtained from the
PHMC algorithm, and $f_A(T/2)$ from the HMC algorithm. 
}}
\end{center}
\end{figure}

A first step for understanding the larger error obtained from the PHMC algorithm
is to look at the distribution of $f_A(T/2)$, which we show in fig.~\ref{fig:fadis}.
It is clearly seen that the distribution as obtained with the PHMC algorithm
spreads much further out, towards large values of $f_A(T/2)$. 
In principle, this effect can come either from a large autocorrelation time
encountered within the PHMC algorithm or from a different sampling in configuration
space. 
To decide on this question, we plot in fig.~\ref{fig:allfa} 
the time evolutions
of various quantities. 

Let us start with 
the correction factor $W$.  
Figure~\ref{fig:allfa}(a) shows that $W$   
can become small, 
assuming values that are clearly much below the average value,
$\langle W \rangle \approx 0.45$. 
At the time when $W\ll 1$, $f_A(T/2)$ assumes very large values as shown in 
fig.~\ref{fig:allfa}(b). 
In fig.~\ref{fig:allfa}(c) we show $W\cdot f_A(T/2)$: the spike in $f_A$ is now suppressed
by the correction factor $W$. 
The PHMC algorithm seems to allow for configurations that make large contributions
to quark correlation functions and partly suppresses these contributions 
in the reweighing procedure through small values
of the (noisy) correction factor. In fig.~\ref{fig:allfa}(d) we show the
Monte Carlo time evolution of $f_A(T/2)$ for the HMC algorithm, 
which looks quite different from that of the PHMC algorithm. 
We conclude that the difference in the variance of
$f_A(T/2)$ is not due to a large autocorrelation time but reflects the fact that
the PHMC algorithm really samples the configuration space differently. 
A similar observation was made in \cite{istvan} in a different context.

\begin{figure}
\vspace{-0mm}
\centerline{ \epsfysize=18.5cm
             \epsfxsize=18.5cm
             \epsfbox{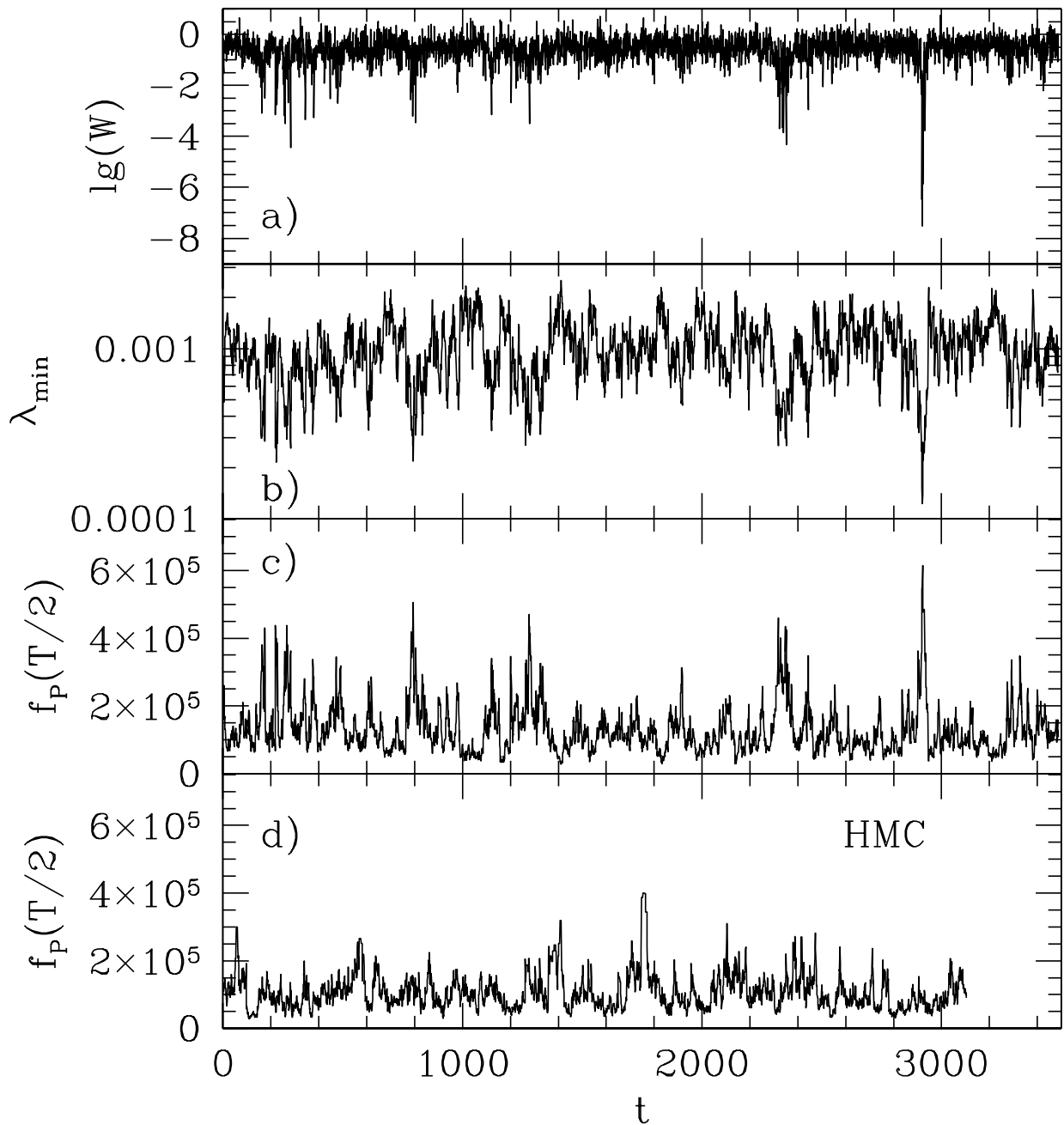}}
\begin{center}
\parbox{12.5cm}{\caption{ \label{fig:allfp}
The Monte Carlo time evolutions for the correction factor $W$, 
$\lambda_{\rm min}(\hat{Q}^2)$, $f_P(T/2)$, as obtained from the
PHMC algorithm, and $f_P(T/2)$ from the HMC algorithm. 
}}
\end{center}
\end{figure}

The Monte Carlo time evolution of $f_P(T/2)$ is plotted in figs.~\ref{fig:allfp}(c,d) 
for the PHMC and the HMC algorithms. For this quantity we do not observe spikes in
the PHMC Monte Carlo history and both time evolutions are similar. 
This is consistent with the results in table 1, where we saw that
the errors for $f_P(T/2)$ are comparable in the two cases.    
In figs.~\ref{fig:allfp}(b,a) we show the time evolution of the lowest eigenvalue
$\lambda_{\rm min}(\hat{Q}^2)$ and $W$ for the PHMC algorithm.

\subsection{Combinations of quark correlation functions} 

Following \cite{paper3,jansom} we define correlation functions 
\begin{eqnarray} \label{rs}
r(x_0) & = & \frac{1}{4}(\partial_0^{\star} + \partial_0) f_A(x_0)/f_P(x_0)
\nonumber \\
s(x_0) & = & \frac{1}{2}\partial_0^{\star}\partial_0 f_P(x_0)/f_P(x_0)
\end{eqnarray}
and analogously $r'(T-x_0)$, $s'(T-x_0)$ in terms of
$f'_A(T-x_0)$ and $f'_P(T-x_0)$. 
These correlation functions allow us to define an                            
unrenormalized PCAC current quark mass: 
\begin{equation} \label{quarkmass} 
M(x_0,y_0) = r(x_0) -s(x_0)\frac{r(y_0)-r'(y_0)}{s(y_0)-s'(y_0)}
\end{equation}
and analogously $M'$.
The non-vanishing of the difference between $M$ and $M'$ at certain time distances 
\begin{equation} \label{deltam}
\Delta M = M(\frac{3}{4}T,\frac{1}{4}T) - M'(\frac{3}{4}T,\frac{1}{4}T)
\end{equation}
is a lattice artefact appearing linear
in the lattice spacing. The requirement that  
$\Delta M$ assumes its tree-level value, $\Delta M =0.000277$,  
is the improvement condition to determine the values of 
$c_{\rm sw}$ non-perturbatively.

\begin{table*}[hbt]
\caption{Comparison of combinations of quark correlation functions as obtained
from both algorithms. Notations are as in table~\ref{tablebulk}. }
\vspace{2mm}
\label{tab:tabledeltam}
\begin{tabular}{lllr}
\hline
Algorithm &  $\tilde{c}_A$ & $M$ & $\Delta M\cdot 10^{3}$ \\                
\hline \hline
 HMC     & $-0.0265(28)[4]$  & $0.00144(36)[5]$ & $0.045(311)[40]$  \\
 PHMC(4) & $-0.0262(30)[4]$  & $0.00161(35)[4]$  & $-0.086(390)[50]$  \\
 PHMC(2) & $-0.0257(26)[3]$  & $0.00155(33)[4]$  & $-0.129(366)[50]$  \\
 PHMC(1) & $-0.0265(31)[4]$  & $0.00150(40)[5]$ & $0.015(381)[50]$  \\
 PHMC(0) & $-0.0242(27)[3]$  & $0.00194(27)[3]$  & $-0.020(493)[60]$  \\
\hline
PHMC$^*$(2) & $-0.0247(56)[7]$  & $0.00180(62)[8]$  & $0.745(404)[50]$  \\
\hline \hline
\end{tabular}
\end{table*}
 
We may build various, physically interesting combinations of
the correlation functions of eqs.~(\ref{rs}). We will consider 
the unrenormalized current quark mass $M$ (eq.~(\ref{quarkmass})), 
$\Delta M$ (eq.~(\ref{deltam})), 
and an {\em estimator} of the improvement coefficient $c_A$, 
\begin{equation}  \label{ca}
 \tilde{c}_A = -\frac{r(T/4)-r'(T/4)}{s(T/4)-s'(T/4)}\; .
\end{equation}
We want to emphasize that $\tilde{c}_A$ should not be taken as 
the true non-perturbatively determined values of $c_A$. We consider
$\tilde{c}_A$ in this work as a purely technical parameter, which
can also be used in comparing the two algorithms. 
We give our results for $M$, $\Delta M$ and $\tilde{c}_A$ in
table~\ref{tab:tabledeltam}. We find that, at least within the statistical
uncertainties, the average values as well as the errors are 
completely compatible. 

We close this section by remarking that we also performed a simulation with the
PHMC algorithm choosing 
a trajectory length of $N_{\rm md}\delta\tau \approx 0.5$. 
The results from this
test are, however, rather inconclusive: while for some observables
the errors did not change with respect to the run with unit trajectory,
for other observables we found an increase of the errors as expected. 

\section{Results at $\beta=5.4$}

This section is devoted to a discussion of the results obtained at
$\beta=5.4$, for which the
lattice spacing $a\approx 0.1$ fm.
We set
$\kappa=0.1379$ and $c_{\rm sw}=1.7275$. 
At these values of the parameters we find a quark mass
$M = 0.009(1)$ \cite{jansom}. 
We use a $8^3\times 16$ lattice and the
boundary conditions are the same as in section 2. 
For reasons that will become clear from our discussion,
we do not aim in this section at a comparison of the HMC and PHMC algorithms
on the same quantitative level as it was done in the previous section
for the results at $\beta=6.8$.
We will rather emphasize the qualitative behaviour of the PHMC algorithm
in sampling configuration space and reweighing observables when
very small eigenvalues of $\hat{Q}^2$ occur.

\subsection{Low-lying eigenvalues}

\begin{figure}
\vspace{-0mm}
\centerline{ \epsfysize=16.5cm
             \epsfxsize=13.5cm
             \epsfbox{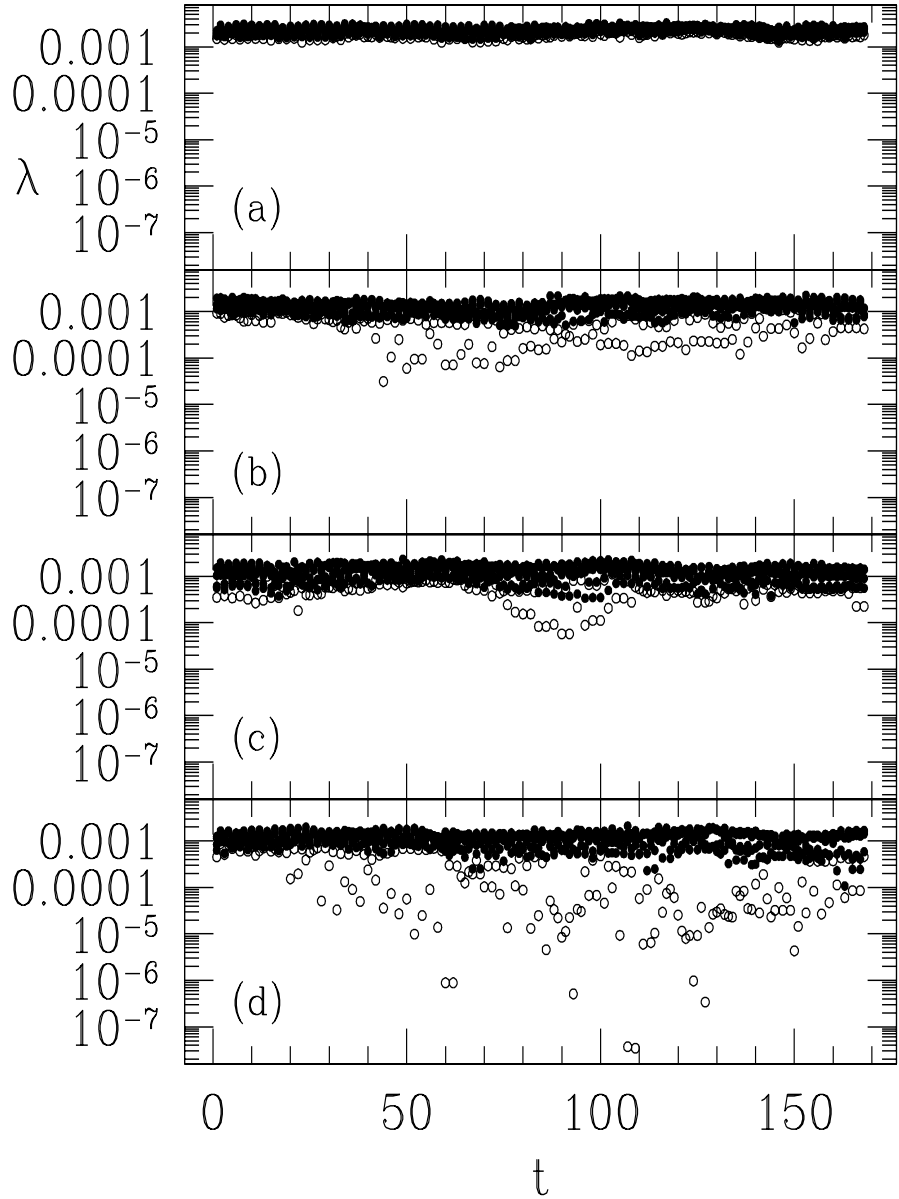}}
\begin{center}
\parbox{12.5cm}{\caption{ \label{fig:evi_4replica}
We show the Monte Carlo time evolution of the five 
lowest eigenvalues of $\hat{Q}^2$ at $\beta=5.4$
in four typical situations. 
The lowest eigenvalue is shown by the open symbols, the remaining eigenvalues 
by the filled ones. 
}}
\end{center}
\end{figure}

As was shown in \cite{hitech} 
for the parameter values 
considered in this section, isolated 
very small eigenvalues of the operator $\hat{Q}^2$ are found. 
We illustrate this again in fig.~\ref{fig:evi_4replica}, by
showing the Monte Carlo time evolution of the five lowest 
eigenvalues in four typical situations.
In fig.~\ref{fig:evi_4replica}(a) the five lowest eigenvalues lie in
a narrow band and we find a 
basically continuous spectrum,
at least up to the tenth eigenvalue. In
figs.\ref{fig:evi_4replica}(b,c) there are a few eigenvalues that assume 
rather small values and finally, in fig.~\ref{fig:evi_4replica}(d), we
observe very small, isolated eigenvalues, lying many orders of magnitude 
below the ones in fig.~\ref{fig:evi_4replica}(a). As can be seen
from the distribution of $\lambda_{\rm min}$ in fig.~3 of ref.~\cite{hitech}, 
such very small eigenvalues could not be observed in the
corresponding simulations using the HMC algorithm.

It is a natural question to ask, whether the occurrence of
the small eigenvalues shown in fig.~\ref{fig:evi_4replica}
is related to some topological effects. We therefore 
consider the values of the
pure gauge action and the naive topological charge
\cite{topo:naive_charge} after performing $500$ cooling \cite{topo:cool} 
iterations (see also \cite{jansom}); these values will be denoted in the following
by $S_{\rm classical}$ and $Q_{\rm topo}$, respectively. 
We emphasize that we do not want to give a precise and reliable
number for the topological charge itself, but rather that we are
interested in the qualitative behaviour of $Q_{\rm topo}$ and 
in only estimating the 
autocorrelation time of a quantity that is related to topology. 
We remark that 
in the case of Schr\"odinger functional boundary conditions
there exist bounds \cite{SF:LNWW} on the pure gauge action $S_g$ given by
\begin{eqnarray} \label{e_act_bound}
 g_0^2 S_g & \geq & \pi^2, \quad Q_{\rm topo}=0 \; , \nonumber \\
 g_0^2 S_g & \geq & 8\pi^2 |Q_{\rm topo}| \; .
\end{eqnarray}

In fig.~\ref{fig:time13} we plot an example of the Monte Carlo time
evolution of $S_{\rm classical}$, $Q_{\rm topo}$ and
the lowest eigenvalue of $\hat{Q}^2$. It is remarkable that, although working
at basically zero quark mass, we see some transitions between
topological sectors. As expected from the bounds of eq.~({\ref{e_act_bound})
the behaviour of $S_{\rm classical}$ closely follows the one of $Q_{\rm topo}$. 

The behaviour of the lowest eigenvalue of $\hat{Q}^2$ and the
topological charge are not as closely related. 
Small eigenvalues are expected when a transition between 
topological sectors occur and, indeed, there is one instance, shown in
fig.~\ref{fig:time13}, where exactly this happens. 
However, we also see from fig.~\ref{fig:time13} that 
the topological charge can change without any occurrence of a very small
eigenvalue. 
This might of course be due to the
fact that a small eigenvalue has appeared during the mo\-le\-cu\-lar dynamics evolution.
Finally we can have situations where the eigenvalue becomes very small
but the topological charge does not change, which might correspond just to an 
unsuccessful attempt to change topological sectors. 
The relation between the topological charge and very small eigenvalues
may be partly obscured in our case by the fact that we use only a naive definition
of the topological charge. Moreover we remark 
that the index theorem does not have to hold owing to the existence of lattice
artefacts and 
to our choice of Schr\"odinger functional boundary conditions.
In any case, since we are 
working at almost zero quark mass and reasonably large physical volume, we take 
fig.~\ref{fig:time13} as an encouraging indication that the PHMC algorithm
is able --even in this physical situation-- to explore different topological 
sectors. Of course, only more extensive investigations, possibly at
larger physical volumes, can decide whether 
our tentative conclusion is too optimistic. 

We remark that when measuring the topological charge with the PHMC algorithm,
its physical value will be the one reweighted with the correction factor. 
If we are close enough to the continuum for the effects of the lattice
spacing to be negligible and we are able to work at vanishing quark mass, 
a non-trivial topological charge has to induce the appearance of a zero mode.
Since the correction factor is proportional to this zero
eigenvalue, the reweighted topological charge will always be zero. 
This corresponds, of course, exactly to the continuum situation,
where the topological charge vanishes after integration over the fermions,      
provided that at least one of the fermion species is massless.

\begin{figure}
\vspace{-0mm}
\centerline{ \epsfysize=16.5cm
             \epsfxsize=13.5cm
             \epsfbox{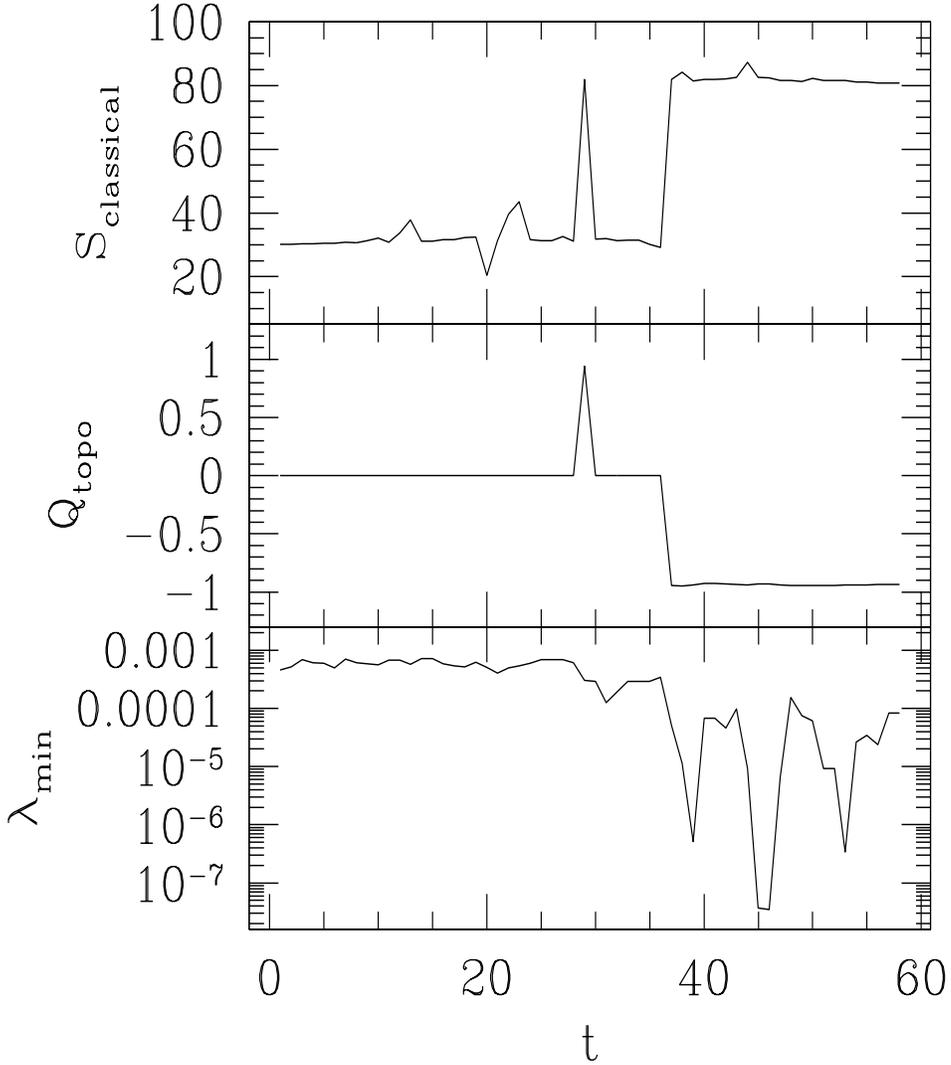}}
\begin{center}
\parbox{12.5cm}{\caption{ \label{fig:time13}
The Monte Carlo time evolution of the pure gauge action and 
the topological charge after cooling, $S_{\rm classical}$ and $Q_{\rm topo}$,
respectively, and the lowest eigenvalue
$\lambda_{\rm min}$, as obtained with the PHMC algorithm at $\beta=5.4$. 
}}
\end{center}
\end{figure}

\subsection{Modified correction factor} 

As discussed in \cite{hitech}, in order to deal with situations where very small
eigenvalues may occur, the correction factor $W$, eq.~(\ref{W_def}), has to
be modified. The reason is that in the presence of very small eigenvalues
the noisy estimate of $\det[ \hat{Q}^2 P_{n,\epsilon}(\hat{Q}^2) ]$ given
in eq.~(\ref{W_def}) is largely dominated by those $\eta$-fields
that are almost orthogonal to all the eigenfunctions of the 
small eigenvalues. Since the probability of extracting such $\eta$-fields
from a distribution $\propto \exp(- \eta^\dagger \eta)$ is low, 
we would need a large value of $N_{\rm corr}$ 
to obtain a good (i.e. not too noisy) 
estimate of $\det[ \hat{Q}^2 P_{n,\epsilon}(\hat{Q}^2) ]$. 
 
An improved definition of the correction factor, replacing eq.~(\ref{W_def}),
is given by (see \cite{hitech} for more details):
\begin{equation} \label{wimproved}
W= W_{\rm B}W_{\rm IR}\; .
\end{equation}
The separation between $W_{\rm B}$ and $W_{\rm IR}$ is controlled by 
a new parameter: $\tilde{\epsilon} \ll \epsilon$. 
In eq.~(\ref{wimproved}) 
$W_{\rm B}$ is a ``bulk'' factor, taking into account the contribution of all
modes with eigenvalues larger than $\tilde{\epsilon}$:
\be \label{W_B_eval}
W_{\rm B}[\eta,U]  = \exp \left\{\eta_{\perp}^\dagger [ R_{n,\epsilon} \cdot
               (\hat{Q}^2\cdot P_{n,\epsilon})^{-1}] \eta_{\perp} \right\} 
\ee
and $W_{\rm IR}$ an ``infrared'' factor that incorporates the 
contribution from the eigenmodes of $\hat{Q}^2$ lying below $\tilde{\epsilon}$,
\be  \label{W_IR_eval}
W_{\rm IR} =  \; \prod_{\lambda_j \le \tilde{\epsilon}} [ 1 + R_{n,\epsilon}(\lambda_j) ] \; .
\ee
In eqs.~(\ref{W_B_eval}) and (\ref{W_IR_eval}) we have introduced
the relative fit deviation $R_{n,\epsilon} = \hat{Q}^2 P_{n,\epsilon} - 1$,
the eigenmodes $| \lambda_j \rangle$ of $\hat{Q}^2$ and the projection
of the $\eta$-field onto the subspace orthogonal to all the modes lying
below $\tilde{\epsilon}$: 
\bea \label{eta_perp}
\hat{Q}^2 | \lambda_j \rangle & = & \lambda_j | \lambda_j \rangle \nn \\
| \eta_{\perp} \rangle & = & | \eta \rangle - \sum_j \theta (\tilde{\epsilon} - \lambda_j)
| \lambda_j \rangle \langle \lambda_j | \eta \rangle \; .
\eea

Whereas $W_{\rm B}$ is given again by a noisy estimator, 
$W_{\rm IR}$ is calculated ``exactly'' in terms of the eigenvalues of $\hat{Q}^2$
that are smaller than $\tilde{\epsilon}$. These eigenvalues
can be explicitly computed, together with
the corresponding eigenvectors, with a pre-defined accuracy \cite{BunkEtAl,KalkreuterSimma}.
In order to guarantee the exactness of the PHMC algorithm,
$\tilde{\epsilon}$ has to be fixed in a simulation beforehand. For the present investigation
we have chosen $\tilde{\epsilon} = \epsilon/10$. 
Clearly, when no eigenvalues smaller than $\tilde{\epsilon}$ occur, $W_{\rm IR}=1$.
In particular, for $\tilde{\epsilon}=0$ we are back to the old correction factor 
of eq.~(\ref{W_def}). 

The difference between the old and the modified correction factors, as
evaluated on a gauge configuration carrying a very small isolated 
eigenvalue ($\lambda_{\rm min} = 3.7 \cdot 10^{-7}$),
is demonstrated
in fig.~\ref{modcorrfact}. There we plot the distribution
for a fixed gauge field configuration 
of $w=\log(W_{\rm B} W_{\rm IR})$ as obtained from the generation of $600$ $\eta$-fields 
for two different values of $\tilde{\epsilon}$. When setting
$\tilde{\epsilon} =0$ (open squares) the distribution is very broad, leading to a
very noisy and imprecise estimate of $\det[ \hat{Q}^2 P_{n,\epsilon}(\hat{Q}^2) ]$. 
When setting
$\tilde{\epsilon} =0.00011$ (filled squares), 
i.e. a value ten times smaller than $\epsilon$,
the distribution appears as a needle on the scale of the upper plot in
fig.~\ref{modcorrfact}. In the lower plot of this figure we therefore resolve
the distribution for $\tilde{\epsilon} =0.00011$. 
The picture nicely demonstrates that 
if we use the original form of the correction factor, i.e. set $\tilde{\epsilon} = 0$,
the estimate of $W[\eta,U]$ is dominated by the terms
$W_{\rm B}[\eta_m,U]$ with $\langle \eta_m | \lambda_{\rm min} \rangle \simeq 0$. However, 
vectors
$|\eta \rangle$, which are almost orthogonal to $ | \lambda_{\rm min} \rangle $, may be extracted
very rarely from the probability distribution $P[\eta] = \exp\{-{\eta^\dagger\eta}\}$. This leads
to large fluctuations affecting the estimate of $\det[ \hat{Q}^2 P_{n,\epsilon}(\hat{Q}^2) ]$,
and a very large value of $N_{\rm corr}$ is needed for the result to be 
sufficiently precise. 
For $\tilde{\epsilon} =0.00011$ 
the distribution becomes much narrower
and a value of $N_{\rm corr}$ lower than $10$ is sufficient to achieve
a precision that is appropriate for the purpose of keeping the fluctuations of $W_{\rm B}$ 
small. 

\begin{figure}
\vspace{-0mm}
\centerline{ \epsfysize=12.5cm
             \epsfxsize=12.5cm
             \epsfbox{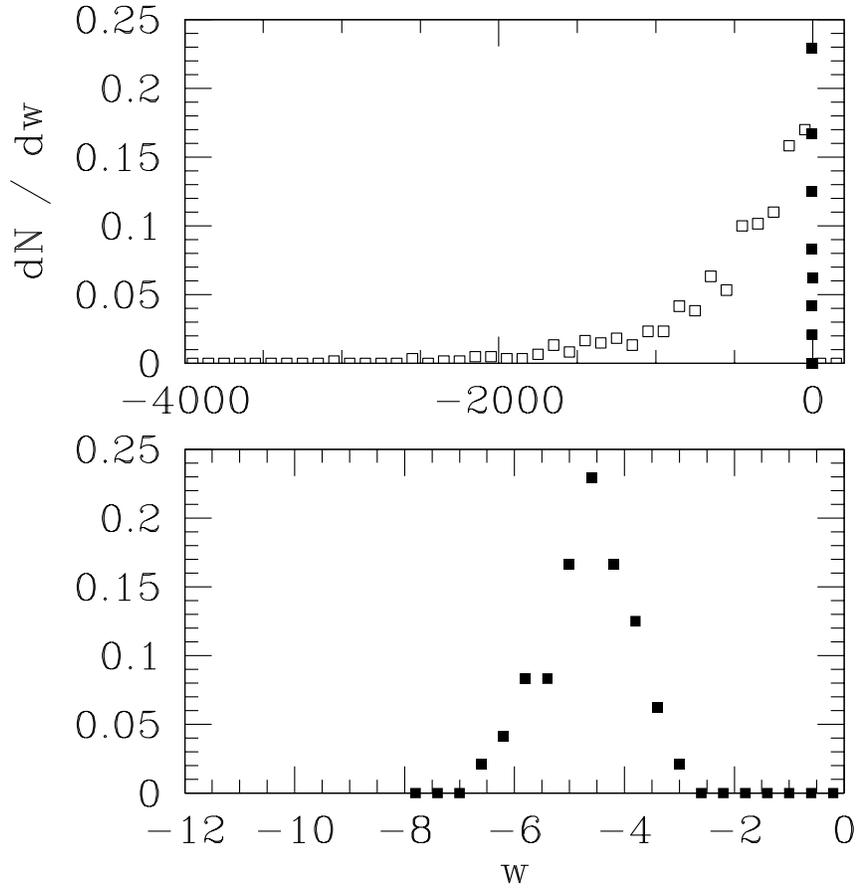}}
\begin{center}
\parbox{12.5cm}{\caption{ \label{modcorrfact}
The distribution of $w=\log(W_{\rm B} W_{\rm IR})$ for $\tilde{\epsilon}=0$ (open squares)
and $\tilde{\epsilon} =0.00011$ (filled squares), on a fixed gauge configuration
carrying an exceptionally small (isolated) eigenvalue of $\hat{Q}^2$. 
In the lower figure, we resolve the distribution of $w$ for the case
$\tilde{\epsilon} =0.00011$. 
}}
\end{center}
\end{figure}

For situations 
where no eigenvalue of $\hat{Q}^2$ is exceptionally
small it should make no difference whether 
$\tilde{\epsilon}$ is set to zero or to some finite value smaller than $\epsilon$.                 
The noise in the estimate of $\det[ \hat{Q}^2 P_{n,\epsilon}(\hat{Q}^2) ]$
will be essentially the same for both cases, since there is no single mode that plays a
dominating role in determining the value of $W_{\rm IR} W_{\rm B}$. We have checked this expectation 
explicitly and our numerical results fully confirm the above picture. 
We have also checked that a relative precision of $1\%$ in the evaluation of the low-lying
eigenvalues of $\hat{Q}^2$ yields eigenvectors that are accurate enough to get a
precision sufficient for the projection onto the subspace orthogonal to the
one spanned by the eigenvectors themselves. 
Concerning the computational cost of 
the modified correction factor, eq.~(\ref{wimproved}), 
an overhead with respect to 
the cost of computing the ordinary correction factor, eq.~(\ref{W_def}), comes
from the evaluation of the needed eigenvalues and eigenvectors of $\hat{Q}^2$.
This overhead depends on the choice of $\tilde{\epsilon}$.           
In our test run at $\beta=5.4$, we found that for a case (see below) when
the four lowest eigenvalues 
and eigenvectors of $\hat{Q}^2$ are needed, the overhead for the modified
correction factor is just
half the time of evaluating the ordinary correction factor
having $N_{\rm corr}=4$. 
We mention that, when setting $\tilde{\epsilon}=\epsilon/10$, 
the modified correction factor had only to be computed 
in about $35\%$ of our measurements. This leads to an additional 
reduction of the overhead.              
We will hence neglect this overhead when discussing computational costs in section 4.

\begin{table*}[hbt]
\caption{The uncorrected ($N_{\rm corr}=0$) values of the ten lowest eigenvalues
of $\hat{Q}^2$: We give the expectation values, with the corresponding true error
in parenthesis, and the variance, as obtained from our PHMC test at $\beta=5.4$,
see table \ref{tech54}. Note that the value of $\epsilon$ was set to $0.0011$. 
Moreover we show the ten lowest eigenvalues of $\hat{Q}^2$ for two particular
gauge configurations (C$_1$ and C$_2$), the first of which has a very small value 
of $\lambda_{\rm min}$.
}
\vspace{2mm}
\label{spectrum_cut}
\begin{tabular}{lllll}
\hline
Eigenvalue & $\langle \lambda\rangle$ 
& $\sqrt{\langle \lambda^2\rangle-\langle \lambda\rangle^2}$ & C$_1$ & C$_2$ \\
\hline \hline
$\lambda_1 = \lambda_{\rm min}$ & $0.00032(5)$ & 0.00024 &  $0.0000017$ &  $0.00052 $   \\
$\lambda_2                    $ & $0.00054(5)$ & 0.00026 &  $0.00027  $ &  $0.00087 $   \\
$\lambda_3                    $ & $0.00090(5)$ & 0.00034 &  $0.00041  $ &  $0.00137 $   \\
$\lambda_4                    $ & $0.00114(6)$ & 0.00032 &  $0.00077  $ &  $0.00152 $   \\
$\lambda_5                    $ & $0.00140(6)$ & 0.00033 &  $0.00098  $ &  $0.00170 $   \\
$\lambda_6                    $ & $0.00162(6)$ & 0.00033 &  $0.00139  $ &  $0.00171 $   \\
$\lambda_7                    $ & $0.00190(6)$ & 0.00032 &  $0.00141  $ &  $0.00189 $   \\
$\lambda_8                    $ & $0.00212(5)$ & 0.00032 &  $0.00204  $ &  $0.00217 $   \\
$\lambda_9                    $ & $0.00237(5)$ & 0.00031 &  $0.00206  $ &  $0.00262 $   \\
$\lambda_{10}                 $ & $0.00256(5)$ & 0.00031 &  $0.00260  $ &  $0.00274 $   \\
\hline \hline
\end{tabular}
\end{table*}
 
In table~\ref{spectrum_cut} we show data for the low end of the spectrum of $\hat{Q}^2$:
for the ten lowest eigenvalues, we consider the
expectation values and the variance of the uncorrected ($N_{\rm corr}=0$)
eigenvalues.
We see that the variance is almost constant and takes a value of the
same order of magnitude as the average lowest eigenvalue of $\hat{Q}^2$.
We also give the example of two particular gauge configurations,
one with an exceptionally small eigenvalue and another with no
exceptional eigenvalues. Note that for the first configuration (C$_1$) all the eigenvalues
$\lambda_j$, with $1<j \leq 10$, lie somewhat below the corresponding
eigenvalues measured for the second configuration (C$_2$).
We infer from the results for the variance that
for practically all gauge configurations of our sample
there are only very few eigenvalues lying below $\epsilon$. This also justifies our
choice of $\tilde{\epsilon} = \epsilon/10$ for the modified correction factor.
We remark that with this choice of $\tilde{\epsilon}$ for evaluating the
modified correction factor, eq.~(\ref{wimproved}), we need not more than the
four lowest modes of $\hat{Q^2}$ (see table~\ref{spectrum_cut}).

Let us finally demonstrate that, despite the different behaviour of the two algorithms
in sampling configuration space, compatible results are found within the present
statistical uncertainties. 
In table~\ref{tech54} we give the algorithmic parameters for the
simulations performed at $\beta=5.4$ as well as the acceptance rates and the statistics.
 
\begin{table*}[hbt]
\caption{Technical parameters for the algorithms at $\beta=5.4$.}
\vspace{2mm}
\label{tech54}
\begin{tabular}{llllllll}
\hline
Algorithm & $\delta\tau$ & $N_{\rm md}$ & $P_{\rm acc}$ & Stat & $\epsilon$ & $n$ & $c_M$  \\
\hline \hline
 HMC  & $0.032$ & $34$ & $0.948(8)$ & $5120$ & --  & -- & -- \\
 PHMC & $0.056$ & $18$ & $0.83(1)$ & $1632$ & $0.0011$ & $76$ & $0.806$  \\
\hline \hline
\end{tabular}
\end{table*}

In table~\ref{tablebulk54} we present a comparison of the bulk quantities.    
Note that the statistics for the HMC run is about a factor of 3 larger.
We emphasize again, however, that with this small statistics the error
on the error is substantial and no real comparison 
of the performance between the two algorithms is possible. 
To really say something about the performance, a much larger statistical sample
would be necessary for both algorithms. Since 
a non-negligible amount of computer time has already been invested
in obtaining the present statistics,
we feel that such a comparison should be made within
a project that aims at the same time at physical results. 

\begin{table*}[hbt]
\caption{Comparison of bulk quantities as obtained from
the HMC and the PHMC algorithms at $\beta=5.4$. Notations are
as in table \ref{tablebulk}.}
\vspace{2mm}
\label{tablebulk54}
\begin{tabular}{lllll}
\hline
Algorithm & \makebox[1.5cm][r]{$\langle P \rangle$}& $\langle \lambda_{\rm min}(\hat{Q}^2) \rangle $
          &   $\langle \lambda_{\rm max}(\hat{Q}^2) \rangle $ & $\langle dS_g/d\eta \rangle$ \\
\hline \hline
 HMC     & $0.563331(65)$ & $0.000561(17)$ & $0.83555(31)$ & $0.8(2.0)$  \\
 PHMC(8) & $0.563302(120)$ & $0.000506(50)$ & $0.83672(99)$ & $-6.2(4.4)$  \\
 PHMC(4) & $0.563344(135)$ & $0.000528(69)$ & $0.83665(90)$ & $-3.6(5.0)$  \\
 PHMC(2) & $0.563404(138)$ & $0.000554(85)$ & $0.83649(190)$ & $-4.6(6.3)$  \\
 PHMC(1) & $0.563679(377)$ & $0.000600(107)$ & $0.83599(259)$ & $-0.7(9.6)$  \\
 PHMC(0) & $0.563336(122)$ & $0.000322(50)$ & $0.83730(43)$ & $-6.9(3.4)$  \\
\hline \hline
\end{tabular}
\end{table*}
 
We remark that we have also monitored the quark correlation functions,
introduced in sections 2.2 and 2.3, finding consistent results for the HMC
and the PHMC algorithm. No new qualitative features arise with respect to
our discussion for the data obtained at $\beta=6.8$
(see the previous section). In particular, we find again 
spikes in the uncorrected quark correlation functions, in coincidence with
gauge configurations carrying very small eigenvalues of $\hat{Q}^2$. For these  
configurations we observe e.g. for $f_A(T/2)$ values up to three orders of magnitudes larger
than the typical values assumed for ``normal'' gauge configurations; at the same time
the modified correction factor, eq.~(\ref{wimproved}), takes values up to
three orders of magnitudes smaller than the usual ones, leading, as expected, to contributions 
of ``normal'' size to the reweighted average, eq.~(\ref{true_ave}).

Taking $N_{\rm corr}=8$ and a statistics of $1632$ trajectories for the
PHMC algorithm, we find for the quark mass $M=0.0066(21)$ and for the lattice
artefact $\Delta M=0.00299(183)$. 
With the same statistics, the HMC algorithm gives
$M=0.0086(28)$ and $\Delta M =0.00026(201)$. 
This indicates, but does not prove, that with the same statistics
compatible errors can be obtained from the two algorithms also for these quantities.

\section{Computational cost}

A crucial question is, of course, how the cost of the PHMC algorithm
compares with the one of the HMC algorithm. 
In this section we will therefore give
the computational cost of both algorithms for generating one gauge field configuration
at the two values of $\beta$ considered in this paper.
For the simulations performed at $\beta=6.8$,  
this comparison of the cost corresponds to a comparison of the actual cost
to generate an independent configuration, because
the errors on almost all observables are compatible between the two algorithms
when the same statistics is employed. 
For the simulation performed at $\beta=5.4$, the situation is, however,
different since, with the available statistics, the uncertainties on the
integrated autocorrelation times are rather large and no definite
statement can be made. 

However, regarding observables for which the very small 
eigenvalues are important, a comparison of the errors would be difficult
even if the statistics were large. 
If the modes corresponding to these very small eigenvalues
are physically important for some observables and the HMC algorithm 
generates these modes only very seldom,
a direct comparison of the fluctuations 
of these particular observables computed with the two algorithms is not appropriate.
This is, of course, a general problem when comparing algorithms
with different behaviour in sampling configuration space 
\footnote{One example would be the behaviour of the cluster
and the Metropolis algorithms at a first-order phase transition.}.
In such a situation the algorithms have very different autocorrelation times. 

In \cite{hitech} we gave a detailed description of the
computational cost of the PHMC algorithm in units of matrix times vector
$Q\phi$ operations. Therefore, we list here only the
formulae for the cost analysis derived in \cite{hitech}. Let us remark that 
the cost of a single trajectory in
both algorithms may be written as 
\begin{equation}
C_{\rm tot} = C_{Q\phi} + C_{\rm extra} \; ,
\end{equation}
where the first contribution is given by the number of matrix
times vector $Q\phi$ operations and the second part accounts for all
other operations.
Asymptotically, when the condition number of $Q$ becomes
large, $C_{Q\phi}$ will by far dominate the cost of the algorithms. We will
therefore only discuss and compare the cost $C_{Q\phi}$ in
the following. 
 
Let us denote by $N_{\rm CG}$ the average number of iterations of the
Conjugate Gradient algorithm that is implemented in our
programs for all matrix inversions\footnote{We remark that in the set up 
we consider here and using APE computers the standard CG solver was found to be competitive 
with the BiCGStab solver for the purpose of inverting $\hat{Q^2}$.}.
Then the cost for the HMC algorithm
in units of $Q\phi$ operations is given by
\begin{equation}
C_{Q\phi}({\rm HMC}) = 2\cdot (2N_{\rm md}+1)\cdot N_{\rm CG}\; ,
\end{equation}
The factor $(2N_{\rm md}+1)$ originates from the use of the
Sexton--Weingarten integration scheme \cite{sexy}.
The cost for the PHMC algorithm is split into three parts \cite{hitech}:
\begin{equation}
C_{Q\phi}({\rm PHMC}) = C_{\rm bhb} + C_{\rm update} + C_{\rm corr}\; ,
\end{equation}
where $C_{\rm bhb}$ is the cost for the heatbath of the bosonic field $\phi$,
$C_{\rm update}$ the cost for the computation of the variation of the
action with respect to the gauge field and $C_{\rm corr}$ the cost
to evaluate the correction factor.
In units of $Q\phi$ operations we find
\begin{eqnarray} \label{PHMC_cost}
C_{\rm bhb} & = & (2n+2)\cdot  N_{\rm CG}^{\rm bhb} + n\nonumber \\
C_{\rm update}  & = & 3n\cdot (2N_{\rm md}+1)  \nonumber \\
C_{\rm corr} & = & (2n+2)\cdot  N_{\rm CG}^{\rm corr} \cdot  N_{\rm corr}\; .
\end{eqnarray}
The factor $N_{\rm corr}$ denotes as usual the number of evaluations
of the correction factor $W$ per full gauge field update (or molecular
dynamics trajectory).
We explicitly verified that the cost in real time, as expected 
from our formulae for $C_{\rm update}$, $C_{\rm bhb}$
and $C_{\rm corr}$, agree with the one measured for our
implementation of the PHMC algorithm on the APE computer.

\begin{table*}[hbt]
\setlength{\tabcolsep}{1.5pc}
\caption{Conjugate Gradient iterations and degree of the polynomial used
in the PHMC runs. Notations are explained in the text.}
\vspace{2mm}
\label{ncg}
\begin{tabular*}
{\textwidth}{@{}l|@{\hspace{1.2cm}}l@{\hspace{1.2cm}}l|l@{\hspace{1.25cm}}l@{\hspace{1.25cm}}l@{\hspace{1.25cm}}}
\hline
                & \multicolumn{2}{c|}{HMC}
                & \multicolumn{3}{c}{PHMC} \\
\cline{1-3} \cline{3-6}
 $\beta$ & $N_{\rm CG}^{32}$ & $N_{\rm CG}$ & $n$ & $N_{\rm CG}^{\rm bhb}$ & $N_{\rm CG}^{\rm corr}$  \\
\hline \hline
 $6.8$ & $149.0(1)$ & $113.3(4)$ & $62$ & $3.66(4)$ & $3.26(4)$  \\
 $6.8$ & --         & --         & $54$ & $3.61(6)$ & $3.88(3)$  \\
 $5.4$ & $197.6(1)$ & $143.0(8)$ & $76$ & $3.56(6)$ & $3.88(4)$  \\
\hline \hline
\end{tabular*}
\end{table*}

All of the simulations done at $\beta=6.8$ and $\beta=5.4$ have been
performed by running several replica in parallel. In particular for the
HMC runs we always had 32 replica. Because the APE computer
we are using is a SIMD machine, all replica have to wait until 
the Conjugate Gradient solver of the slowest replicum has converged. 
This ``parallelization effect'' has an important consequence for the HMC algorithm. 
We give in table~\ref{ncg} the maximal
number of CG iterations, $N_{\rm CG}^{32}$, as determined from the slowest
replicum and the number of CG iterations $N_{\rm CG}$, obtained by averaging over all
replica. As we see from the tables, in particular for $\beta=5.4$, there
can be a substantial increase of the number of CG iterations from
this parallelization effect. The analogous effect is much less relevant in the
case of the PHMC algorithm, since it may occur only in $C_{\rm bhb}$ and $C_{\rm corr}$,
which are asymptotically marginal in comparison with $C_{\rm update}$. To be conservative
in the estimate of the computational cost for the PHMC algorithm, we will neglect to
correct for this small parallelization effect. We do mention, however, that doing so may 
reduce the values for $C_{\rm bhb}$ and $C_{\rm corr}$ by a factor of 2 at $\beta=5.4$.

From tables~\ref{tech68}, \ref{tech54} and \ref{ncg} we
can now calculate the computational cost for both algorithms. We present the results in 
table~\ref{cost68} for $\beta=6.8$ and in table~\ref{cost54} for $\beta=5.4$. 
We give the global costs for both algorithms considering the case of
32 replica ($C_{Q\phi}^{32}$), where the HMC algorithm is slowed down by a significant
parallelization effect, and the case of a single lattice system ($C_{Q\phi}$).

\begin{table*}[hbt]
\setlength{\tabcolsep}{1.5pc}
\caption{ Computational cost for $\beta=6.8$. We take $N_{\rm corr}=1$ for the
PHMC run with $n=62$ (PHMC) and $N_{\rm corr}=2$ for the PHMC run with $n=54$ (PHMC$^{*}$).
The cost $C_{Q\phi}^{32}$ takes the parallelization effect into account when
running 32 replica in parallel. $C_{Q\phi}$ would be the cost when simulating a single
lattice system. }   
\vspace{2mm}
\label{cost68}
\begin{tabular*}{\textwidth}{lllllll}
\hline
 Algorithm & $C_{\rm bhb}$ & $C_{\rm update}$ & $C_{\rm corr}$ & $C_{Q\phi}^{32}$ & $C_{Q\phi}$ \\
\hline \hline
  HMC       &   ---    & $10192$ & ---   & $10192$ & $7750$ \\
 PHMC       & $523$    & $5022$  & $411$ & $5956$  & $5956$ \\
PHMC$^{*}$  & $451$    & $4374$  & $854$ & $5679$  & $5679$ \\
\hline
\end{tabular*}
\end{table*}

\begin{table*}[hbt]
\setlength{\tabcolsep}{1.5pc}
\caption{Pure computational cost for $\beta=5.4$. We consider the cases $N_{\rm corr}=4$
(PHMC(4)) and $N_{\rm corr}=8$ (PHMC(8)).
Notations are as in table~\ref{cost68}.}   
\vspace{2mm}
\label{cost54}
\begin{tabular*}{\textwidth}{lllllll}
\hline
 Algorithm & $C_{\rm bhb}$ & $C_{\rm update}$ & $C_{\rm corr}$ & $C_{Q\phi}^{32}$ & $C_{Q\phi}$ \\
\hline \hline
  HMC       &   ---    & $27269$ & ---   & $27269$ & $19734$ \\
 PHMC(4)       & $624$    & $8436$  & $2390$ & $11450$  & $11450$ \\
 PHMC(8)       & $624$    & $8436$  & $4780$ & $13840$  & $13840$ \\
\hline
\end{tabular*}
\end{table*}
 
For $\beta=6.8$ we see that the dominating effect in the cost gain of the PHMC algorithm
stems from the parallelization effect. Taking this effect out, we still have a 
performance of the PHMC algorithm better than that of the 
HMC algorithm, but the gain becomes marginal.                    
We remark that at $\beta=6.8$ the lattice spacing is very small and we are hence working 
in a correspondingly small physical volume. 
Going to a more challenging situation, i.e. $\beta=5.4$,
we still find a large parallelization effect but now even if this is taken out,
a factor of almost 2 is found in favour of the PHMC algorithm. 
We emphasize again at this point that we give here only the computational
cost of the algorithms and do not take the autocorrelation time into account
for the reasons discussed above.

\section{Conclusions}

In this paper we have tested the PHMC algorithm for
${\rm O}(a)$-improved Wilson fermions. We compared 
the computational cost of the PHMC algorithm, as well as its
qualitative behaviour, with those of the HMC algorithm. 
Practical simulations were performed on $8^3\times 16$ lattices
at $\beta=6.8$, which corresponds to a very small physical volume,
and $\beta=5.4$, corresponding to an intermediate physical volume,
with a lattice spacing $a\approx0.1$ fm. The results of our tests
lead us to the following conclusions:

\begin{itemize}
\item [1)] It is {\em easy} to find values for the degree $n$ and the infrared 
           parameter $\epsilon$, determining 
           the polynomial approximation used in the PHMC algorithm, such that its
           performance becomes comparable to that of the HMC algorithm. 
           As a guideline one may choose $\epsilon \approx 2\langle\lambda_{\rm min}\rangle$,
           with $\lambda_{\rm min}$ the lowest eigenvalue of the fermion matrix
           used in the simulation. The degree $n$ of the polynomial should then
           be chosen such that $\delta \leq 0.01$, see eq.~(\ref{delta}). 

\item [2)] With some extra tuning of $n$ and $\epsilon$ it is possible 
           to improve on the computational cost of the PHMC algorithm and 
           a gain over the HMC algorithm can be obtained that can
           reach about a factor of 2. In particular it seems that  
           when going to larger physical volumes this gain tends to increase.
           Another --substantial-- gain can be obtained from the PHMC algorithm
           on massively parallel machines 
           when several replica are run in parallel. 

\item [3)] Even if one decides to conservatively choose 
           the polynomial parameters $n$ and $\epsilon$, 
           such that the computational cost becomes comparable to the one of the
           HMC algorithm,  
           we still see a conceptual 
           advantage of the PHMC algorithm. It samples configuration space differently
           from the HMC algorithm, 
           allowing in particular for exceptionally small eigenvalues 
           of the lattice Dirac operator to occur. 
           Fermionic observables that are proportional to the inverse of these eigenvalues
           get corrected by the correction factor which makes the PHMC algorithm
           exact, yielding a finite contribution to the (reweighted) sample average. 
           We demonstrated this feature in a number of tests
           in this paper and showed that our way of treating these exceptional
           eigenvalues in the simulation is working in practise. 
           If gauge configurations, carrying exceptionally small eigenvalues, 
           are physically important for some observables, the HMC algorithm,
           given its difficulty to generate such configurations, 
           would have a very long 
           autocorrelation time for these quantities. 
           In this scenario the performance gain of the PHMC algorithm would be
           very large. Of course, an investigation of this issue
           is very expensive and should be performed --in our opinion-- 
           within projects aiming at the same time at physical results.

\end{itemize}

\vspace{0.5cm}
{\large\bf Acknowledgements}
 
This work is part of the ALPHA collaboration research programme.
We are most grateful to S. Aoki, A.D. Kennedy, I. Montvay, 
R. Sommer, P. Weisz and U. Wolff
for many useful discussions and helpful comments.
In particular we thank 
M. L\"uscher for essential advice and discussions.
We also thank
DESY for allocating computer time to this project. R.F. thanks the
Alexander von Humboldt Foundation for the financial support for his
research stay at DESY--Hamburg, where part of this work was done.

\end{document}